\documentclass[12pt,a4paper]{article}

\usepackage{a4wide}
\usepackage{amsmath}
\usepackage{bm}
\usepackage{amssymb}
\usepackage{hyperref}
\usepackage{epic}

\newcommand{\Op}{\mathcal{O}}

\newcommand{\zt}{\zeta_3}
\newcommand{\zfr}{\zeta_4}
\newcommand{\zf}{\zeta_5}

\newcommand{\CA}{{C_A}}
\newcommand{\CF}{{C_F}}
\newcommand{\nf}{{n_f}}
\newcommand{\TF}{{T_F}}
\newcommand{\NF}{{N_F}}

\newcommand{\als}{{a_s}}

\newcommand{\pslash}{p \! \! \! /}

\begin{document}

\thispagestyle{empty}


\begin{center}
{\Large{\bf
Four loop anomalous dimension of the second~moment \\[3mm]
of the non-singlet twist-2 operator in QCD
}}
\vspace{15mm}

{\sc
V.~N.~Velizhanin}\\[5mm]

{\it Theoretical Physics Department\\
Petersburg Nuclear Physics Institute\\
Orlova Roscha, Gatchina\\
188300 St.~Petersburg, Russia}\\[7mm]

\textbf{Abstract}\\[2mm]
\end{center}

\noindent{
We present the result of a calculation for the first even moment of the non-singlet four-loop
anomalous dimension of Wilson twist-2 operators in QCD with full color and flavor structures.
}
\newpage

\setcounter{page}{1}

Calculation of anomalous dimensions of the Wilson twist-2 operators is one of the part of
operator product expansion for the structure functions
in the framework of perturbative Quantum Chromodynamics (QCD).
At the present time such calculations are performed up to three-loop
order~\cite{Gross:1973ju,Georgi:1951sr,Floratos:1977au,GonzalezArroyo:1979df,Larin:1991fx,Larin:1993vu,Moch:2004pa},
while other part of operator product expansion, the coefficient functions, which are known in
the same order \cite{Bardeen:1978yd,Larin:1991fv,Larin:1996wd,Vermaseren:2005qc}, demand the
four-loop anomalous dimensions.

In this paper we present the result of calculations for the first even moment of the
non-singlet anomalous dimension of Wilson twist-2 operators at fourth order in perturbative
QCD.
Similar result can be found in Ref.~\cite{Baikov:2006ai}, but our result contains full color
and flavor structures and the calculations are performed with a different
method\footnote{Note, that there is all-loop prediction for the ${\mathcal O}(1/N_f)$
contribution to the non-singlet anomalous dimension of twist-2 operators in
QCD~\cite{Gracey:1994nn}.}.

The moments of the structure functions $F_k$ are expressed through the parameters of the
following operator product expansion of the $T$-product of electromagnetic currents:
\begin{eqnarray}
T_{\mu\nu}&=&i \int d^4 z\ e^{iqz} T\left\{J_\mu(z)J_\nu(0)\right\}\nonumber\\
&=&\sum\Bigg[\left(g_{\mu\nu}-\frac{q_\mu q_\nu}{q^2}\right)
q_{\mu_1}q_{\mu_2}C^a_{L,N}\!\left(\frac{Q^2}{\mu^2},a_s\right)\nonumber\\
&&-\Big(g_{\mu\mu_1}g_{\nu\mu_2}q^2
      -g_{\mu\mu_1}q_{\nu}q_{\mu_2}
      -g_{\nu\mu_2}q_{\mu}q_{\mu_1}
      -g_{\mu\nu}q_{\mu_1}q_{\mu_2}\Big)
C^a_{2,N}\!\left(\frac{Q^2}{\mu^2},a_s\right)\Bigg]\nonumber\\
&&\times\ q_{\mu_3}\ldots q_{\mu_N}\left(\frac{1}{Q^2}\right)^N
\Op^{a,\{\mu_1\ldots\mu_N\}}\nonumber\\
&&+\ \mbox{singlet contributions} + \mbox{higher twists}\,, \label{OPE}
\end{eqnarray}
where it is usual to use the following notation
\begin{equation}
a_s=\frac{g^2}{16\pi^2}=\frac{\alpha_s}{4\pi}
\end{equation}
for the QCD strong coupling constant. The sum in eq. (\ref{OPE}) runs over the standard set of
the spin-$N$, twist-2 irreducible (i.e. symmetrical and traceless in indices
$\mu_1\ldots\mu_N$) flavor non-singlet quark operators:
\begin{equation}
\Op^{a,\{\mu_1\ldots\mu_N\}}=\bar\psi\lambda^a
\gamma^{\{\mu_1}{\mathcal D}^{\mu_2}\ldots {\mathcal D}^{\mu_N\}}\psi,\qquad a=1,2,\ldots,8,
\label{NSOpN}
\end{equation}
where ${\mathcal D}^{\mu_j}$ are the covariant derivatives, $\lambda^a$ are the generators of
the flavor group $SU(n_f)$ and $C^a_{k,N}(Q^2/\mu^2,a_s)$ are the corresponding coefficient
functions.

The non-singlet moments of the structure functions $F_k$ are expressed through operator
product expansion (\ref{OPE}) in the following form:
\begin{eqnarray}
M_{k,N}&=&\int d^x x^{N-2}F_k^{ep-en}(x,Q^2)\nonumber\\
&&=\sum_a
C^a_{k,N}\!\left(\frac{Q^2}{\mu^2},a_s\right)
\left[A^a_{N,\mathrm{proton}}(\mu^2)-A^a_{N,\mathrm{neutron}}(\mu^2)\right] \label{MomentSF}
\end{eqnarray}
where $A_{N,\mathrm{nucleon}}$ is the spin-averaged nucleon matrix elements of the operator:
\begin{equation}
\langle p,\mathrm{nucleon}|\Op^{a,\{\mu_1\ldots\mu_N\}}|\mathrm{nucleon},p\rangle=
p^{\{\mu_1}\ldots p^{\mu_N\}}
A^a_{N,\mathrm{nucleon}}(\mu^2)\label{MatrixElement}\,.
\end{equation}

Application of the renormalization group technique gives for the coefficient functions the
following standard expression:
\begin{equation}\label{CF}
C^a_{k,N}\!\left(\frac{Q^2}{\mu^2},a_s(\mu^2)\right)=
C^a_{k,N}\!\left(1,a_s(Q^2)\right)
\exp\!\left(-\int_{a_s(\mu^2)}^{a_s(Q^2)}d a'_s\frac{\gamma(a'_s)}{\beta(a'_s)}\right).
\end{equation}
The anomalous dimensions $\gamma_N$ in eq. (\ref{CF}) are defined as
\begin{equation}\label{ADdef}
\gamma_N(a_s)=\mu^2\frac{d\,\log Z_N}{d\,\mu^2}=\sum_{n=0}^{\infty}\gamma_N^{(n)}a_s^{(n+1)}
\end{equation}
and renormalized operators and bare ones are connected as follows:
\begin{equation}
\left(\Op^{a,\{\mu_1\ldots\mu_N\}}\right)_R=
(Z_N)^{-1}\left(\Op^{a,\{\mu_1\ldots\mu_N\}}\right)_B\label{RenormOp}.
\end{equation}
The four-loop approximation for the $\beta$-function in QCD in the MS-scheme was obtained in
Refs.~\cite{Tarasov:1980au,vanRitbergen:1997va}:
\begin{eqnarray}
\beta(a_s)&=&-\beta_0 a_s^2-\beta_1 a_s^3-\beta_2 a_s^4-\beta_3
a_s^5+O\left(a_s^6\right)\,,\\
\beta_0 &=& 11-\frac{2}{3}n_f\,,\\
\beta_1 &=& 102 - \frac{38}{3}n_f\,,\\
\beta_2 &=& \frac{2857}{2}-\frac{5033}{18}n_f +\frac{325}{54}n^2_f\,,\\
\beta_3 &=& \left(\frac{149753}{6}+ 3564\,\zeta_3\right)
-\left(\frac{1078361}{162}+\frac{6508}{27}\,\zeta_3\right)n_f\nonumber\\
&&+\left(\frac{50065}{162}+\frac{6472}{81}\,\zeta_3\right)n^2_f
+\frac{1093}{729}n^3_f\,.
\end{eqnarray}

The perturbative expansion in $a_s$ for the coefficient functions is
\begin{equation}
C_{k,N}(1, a_s)= B_{k,N}^{(0)}+
B_{k,N}^{(1)}a_s+
B_{k,N}^{(2)}a_s^2+
B_{k,N}^{(3)}a_s^3+
O\left(a_s^4\right),
\label{CFexpansion}
\end{equation}
where the Callan—Gross relation gives $B_{L,N}^{(0)}=0$ for all N and the standard deep
inelastic normalization \cite{Bardeen:1978yd} of the coefficient functions implies $B_{2,N}^{(0)}=1$.

The next-next-next-to-leading (NNNL) approximations for the non-singlet moments after
renormalization group improvement are
\begin{equation}
M_{2,N}(Q^2) = a_s^{\gamma_N^{(0)}/\beta_0}
\Big(
B_{2,N}^{(0)}
+B_{2,N}^{(1)}a_s
+B_{2,N}^{(2)}a_s^2
+B_{2,N}^{(3)}a_s^3
\Big)
E(a_s)A_N(\mu^2)
\label{MomentExpansion}
\end{equation}
with
\begin{eqnarray}
E(a_s)& =& 1
+\frac{a_s}{\beta _0^2}E_1
+\frac{a_s^2}{\beta _0^4}
\left[
\frac{E_1^2}{2}
-E_1 \beta _0 \beta _1
+E_2 \beta _0^2
\right]\nonumber\\
&&+\,\frac{a_s^3}{ \beta _0^6}
\left[\frac{E_1^3}{6}
- E_1^2 \beta _1 \beta _0
+ E_1\Big( \left(\beta _1^2-\beta _0 \beta _2\right) \beta _0^2
+E_2 \beta _0^2\Big)
- E_2 \beta _1 \beta _0^3
+ E_3 \beta _0^4
\right]\!,
   \end{eqnarray}
where $E_i=\gamma_N^{(i)}\beta_0-\gamma_N^{(0)}\beta_i$ and
for the NNNL approximation one should keep the four leading orders in the product of the power
series for the coefficient functions with the series
$E(a_s)$.
So, for the NNNL approximation to the non-singlet moments $M_{2,N}$ we need to known the
3-loop coefficients $B_{2,N}^{(3)}$ for the coefficient functions $C_{2,N}(1, a_s)$ and the
4-loop coefficients $\gamma_N^{(3)}$ for the anomalous dimensions $\gamma_N(a_s)$.

For the first even moment ($N=2$) non-singlet operator (\ref{NSOpN}) has the following form
\begin{equation}
{\mathcal O}^{a,\{\mu\nu\}}_{\mathrm {NS}}
\ =\ \bar{\psi} \lambda^a\gamma^\mu {\mathcal D}^\nu\psi\
+\ \bar{\psi} \lambda^a\gamma^\nu {\mathcal D}^\mu \psi\
-\ \frac{2}{D} g^{\mu\nu}\bar{\psi} \lambda^a\gamma^\sigma {\mathcal D}_\sigma
\psi\,,\label{NSOp2}
\end{equation}
where $D=4-2\epsilon$ is space-time dimension.

The calculation of the anomalous dimension of such operators can be performed in a usual way
through the computation
of the Green's function with the operator insertion, which have the following general form in
momentum space (see \cite{Gracey:2003mr}):
\begin{eqnarray}
{\mathcal G}^{\mu\nu}_{{\mathcal O}_{\mathrm {NS}}} (p) &=&
\langle \psi(p) \ [\, {\mathcal O}^{a,\{\mu\nu\}}\,](0) \ \bar{\psi}(-p) \rangle \nonumber \\
&=& \Sigma^{(1)}_{{\mathcal O}_{\mathrm {NS}}}(p) \left( \gamma^\mu p^\nu
+ \gamma^\nu p^\mu - \frac{2}{D} \pslash g^{\mu\nu} \right) \nonumber \\
&& +~ \Sigma^{(2)}_{{\mathcal O}_{\mathrm {NS}}}(p) \left( p^\mu p^\nu
\pslash - \frac{p^2}{D} \pslash g^{\mu\nu} \right)\,,\label{Green2}
\end{eqnarray}
where $p$ is the momentum flowing through the external quark legs.
To determine different components we use the following projectors (see \cite{Gracey:2003mr})
\begin{eqnarray}
\textbf{P}^{(1)}_{{\mathcal O}_{\mathrm {NS}}}(p) &=& \frac{1}{8(D-1)} \left[
\mbox{tr}\!
\left(
\gamma_\mu p_\nu + \gamma_\nu p_\mu - \frac{2}{D}
\pslash g_{\mu\nu}
\right)
-\ 2 \, \mbox{tr}\!
\left( p_\mu p_\nu \pslash
- \frac{p^2}{D} \pslash g_{\mu\nu} \right)
\right],\label{Projector1}
\\
\textbf{P}^{(2)}_{{\mathcal O}_{\mathrm {NS}}}(p) &=&
\frac{-1}{4(D-1)} \left[ \mbox{tr}\!
\left(
\gamma_\mu p_\nu + \gamma_\nu p_\mu
- \frac{2}{D} \pslash g_{\mu\nu}
\right)
- (D+2) \, \mbox{tr}\!
\left(
p_\mu p_\nu
\pslash - \frac{p^2}{D} \pslash g_{\mu\nu}
\right)
\right]\!.
\end{eqnarray}

Really, to find the anomalous dimension of the operator ${{\mathcal O}_{\mathrm {NS}}}$ we
should compute only  $\Sigma^{(1)}_{{\mathcal O}_{\mathrm {NS}}}(p)$.
A total number of four-loop diagrams is 12816. As in our previous
work~\cite{Velizhanin:2008jd,Velizhanin:2008pc,Velizhanin:2009gv,Velizhanin:2010vw,Velizhanin:2010ey}
all calculations were performed with FORM~\cite{Vermaseren:2000nd}, using FORM package
COLOR~\cite{vanRitbergen:1998pn} for evaluation of the color traces.
For the dealing with a huge number of diagrams we use a program DIANA~\cite{Tentyukov:1999is},
which call QGRAF~\cite{Nogueira:1991ex} to generate all diagrams.
For evaluation of Feynman integrals we used the method from
Refs.~\cite{Misiak:1994zw,Chetyrkin:1997fm} and our own implementation of the Laporta's
algorithm~\cite{Laporta:2001dd} in the form of the MATHEMATICA package BAMBA with the master
integrals from Ref.~\cite{Czakon:2004bu}.

For the renormalization we need the three-loop renormalization constant for the operator
insertion $g\bar{\psi}\lambda^a \gamma^{\{\mu} {\mathcal A}^{\nu\}} \psi$ with two quarks and
one gluon legs, which can be obtained order by order in a usual way from the renormalization
of operator $
{\mathcal O}^{a,\{\mu\nu\}}_{\mathrm {NS}}$ as
\begin{equation}
Z_{\bar{\psi}\lambda^a \gamma^{\{\mu} {\mathcal A}^{\nu\}} \psi}=
Z_{{\mathcal O}^{a,\{\mu\nu\}}_{\mathrm {NS}}}Z_{{\mathcal A}}^{1/2}Z_g^{1/2}Z_\psi\,,
\end{equation}
where $Z_{{\mathcal A}}$, $Z_g$ and $Z_\psi$ are the renormalization constants for gluon filed
${\mathcal A}^\mu$, coupling constant and quark correspondingly.

Our final result is
\begin{eqnarray}
\gamma^{4-loop}_{{\mathrm{NS}}}(2)&=&
\als\frac{8}{3} \CF
+\als^2 \left[
\frac{376}{27}\CF \CA
-\frac{128}{27} \CF \nf \TF
-\frac{112 }{27}\CF^2
\right]\nonumber\\
&&\hspace*{-14mm}+\,\als^3
\bigg[
\CF^3 \left(\frac{128}{3}\zt-\frac{560}{243}\right)
+\CF^2 \TF \nf \left(\frac{128}{3} \zt-\frac{6824}{243}\right)
+\CF^2 \CA \left(-64 \zt-\frac{8528}{243}\right)\nonumber\\
&&\hspace*{-10mm}\qquad
-\frac{896}{243}\CF \TF^2 \nf^2
+\CF \CA \TF \nf \left(-\frac{128}{3} \zt-\frac{6256}{243}\right)
+\CF \CA^2 \left(\frac{64}{3} \zt+\frac{20920}{243}\right)
\bigg]\nonumber\\&&\hspace*{-14mm}
+\,\als^4 \bigg[
\CF^4 \left(\frac{10880}{81} \zt-\frac{1280 }{3}\zf+\frac{194392}{2187}\right)
\nonumber\\&&\hspace*{-10mm}\qquad
+\CF^3 \TF \nf \left(-\frac{5056 }{81} \zt+\frac{256 }{3}\zfr-\frac{1280
}{3}\zf+\frac{381824}{2187}\right)
\nonumber\\&&\hspace*{-10mm}\qquad
+\CF^3 \CA \left(\frac{31040}{81} \zt-\frac{704 }{3}\zfr+\frac{1280
}{3}\zf+\frac{238676}{2187}\right)
\nonumber\\&&\hspace*{-10mm}\qquad
+\CF^2 \TF^2 \nf^2 \left(-\frac{512 }{3} \zt+\frac{256 }{3}\zfr+\frac{99776}{2187}\right)
\nonumber\\&&\hspace*{-10mm}\qquad
+\CF^2 \CA \TF \nf \left(\frac{25856 }{27} \zt-\frac{1088 }{3}\zfr+\frac{640
}{9}\zf-\frac{355496}{2187}\right)
\nonumber\\&&\hspace*{-10mm}\qquad
+\CF^2 \CA^2 \left(-\frac{25744 }{27} \zt+352 \zfr+\frac{4480
}{9}\zf-\frac{1626064}{2187}\right)
\nonumber\\&&\hspace*{-10mm}\qquad
+\CF \TF^3 \nf^3 \left(\frac{1024 }{81} \zt-\frac{8192}{2187}\right)
+\CF \CA \TF^2 \nf^2 \left(\frac{512 }{3} \zt-\frac{256 }{3}\zfr+\frac{25400}{729}\right)
\nonumber\\&&\hspace*{-10mm}\qquad
+\CF \CA^2 \TF \nf \left(-\frac{8080 }{9} \zt+\frac{832 }{3}\zfr+\frac{8960
}{27}\zf-\frac{106036}{243}\right)
\nonumber\\&&\hspace*{-10mm}\qquad
+\CF \CA^3 \left(\frac{34936 }{81} \zt-\frac{352 }{3}\zfr-\frac{12160
}{27}\zf+\frac{1734130}{2187}\right)
\nonumber\\&&\hspace*{-10mm}\qquad
+\frac{32}{9}\frac{ d_F^{abcd}d_A^{abcd}}{\NF}(62 \zt+160 \zf-23)
+\nf\frac{128}{9}\frac{d_F^{abcd}d_F^{abcd}}{\NF}(16 \zt-40 \zf+13)
\bigg],\label{ADM2}
\end{eqnarray}
where (see Ref.~\cite{vanRitbergen:1997va})
\begin{eqnarray}
\frac{d_F^{abcd}d_F^{abcd}}{N_A}&\ =\ &\frac{N_c^4-6N_c^2+18}{96N_c^2}\,,\label{d44FF}\\
\frac{d_F^{abcd}d_A^{abcd}}{N_A}&\ =\ &\frac{N_c(N_c^2+6)}{8}\,,\label{d44FA}\\
\frac{d_A^{abcd}d_A^{abcd}}{N_A}&\ =\ &\frac{N_c^2(N_c^2+36)}{24}\label{d44AA}
\end{eqnarray}
and for the color group $SU(N_c)$ the more simple Casimir operators are:
\begin{equation}
T_F=\frac12\,,\qquad
C_F=\frac{N_c^2-1}{2N_c}\,,\qquad
C_A=N_c\,,\qquad
N_A=N_c^2-1\,.
\end{equation}
The obtained result coincides with the existing result for the first even moment of the
four-loop non-singlet anomalous dimension from Ref.~\cite{Baikov:2006ai} if we substitute the
explicit expression for all Casimir operators for QCD with three active quarks (i.e. for the
gauge group $SU(3)$ with $n_f=3$).
Moreover, the part of our result, which is proportional to $(n_f)^{i-1}a_s^i$, coincide with
the prediction from Ref.~\cite{Gracey:1994nn}, while the non-planar part was calculated by us
in Ref.~\cite{Velizhanin:2010ey}.

The substitution of the colour factors with $N_c = 3$ into eq.(\ref{ADM2}) gives the following
result for QCD case
\begin{eqnarray}
\gamma^{4-loop}_{{\mathrm{NS}}}(2)&=&
+\frac{32 \als}{9}
+\als^2 \left(\frac{11744}{243}-\frac{256 n_f}{81}\right)\nonumber\\
&&+\als^3 \left(-\frac{896 n_f^2}{729}+n_f \left(-\frac{1280
\zt}{27}-\frac{167200}{2187}\right)+\frac{1280
\zt}{81}+\frac{5514208}{6561}\right)\nonumber\\
&&
+\als^4 \Bigg(
\frac{26060864 \zt}{6561}-\frac{7040
\zfr}{27}-\frac{1249280\zf}{243}+\frac{3100369144}{177147}\nonumber\\
&&\quad
+n_f \left(-\frac{6322976\zt}{2187}+\frac{64640 \zfr}{81}+\frac{14720
\zf}{9}-\frac{167219672}{59049}\right)
\nonumber\\
&&\quad
+n_f^2 \left(\frac{2560 \zt}{27}-\frac{1280 \zfr}{27}+\frac{1084904}{19683}\right)
+n_f^3 \left(\frac{512 \zt}{243}-\frac{4096}{6561}\right)
\Bigg)\,.
\end{eqnarray}

In conclusion we give the explicit results for the different number of active quarks:
\begin{eqnarray}
\gamma^{4-loop}_{{\mathrm{NS}}}(2,n_f=3) &=& 3.55556 \,\als + 38.84774 \,\als^2 + 448.07162
\,\als^3 + 6532.13656 \,\als^4\,, \\
\gamma^{4-loop}_{{\mathrm{NS}}}(2,n_f=4) &=& 3.55556 \,\als + 35.68724 \,\als^2 + 306.02989
\,\als^3 + 3679.66906 \,\als^4\,, \\
\gamma^{4-loop}_{{\mathrm{NS}}}(2,n_f=5) &=& 3.55556 \,\als + 32.52675 \,\als^2 + 161.53001
\,\als^3 + 1108.56696 \,\als^4\,, \\
\gamma^{4-loop}_{{\mathrm{NS}}}(2,n_f=6) &=& 3.55556 \,\als + 29.36626 \,\als^2 + 14.571953
\,\als^3 - 1169.71912 \,\als^4\,.
\end{eqnarray}

It is interesting to compare our result with the predictions~\cite{Kataev:1999bp}\footnote{We thank A.L.~Kataev, who pointed out to us this result}, coming from
the Pade resummation, which for $\gamma^{(3)}_{{\mathrm{NS}}}(2,n_f=4)$ gives $2629$ or $2557$ depending on the resummations procedure. Note, that in four-loop order new colour structures (\ref{d44FF})-(\ref{d44AA}) appear, which can disimprove resummation. So, we give below our result for $\gamma^{(3)}_{{\mathrm{NS}}}(2)$ with the contributions from different colour structures:
\begin{eqnarray}
\gamma^{(3)}_{{\mathrm{NS}}}(2,n_f=3) &=& 4626.76262 + 1932.76417\, d_{44}^{FA} - 27.39022 \, d_{44}^{FF}\,, \\
\gamma^{(3)}_{{\mathrm{NS}}}(2,n_f=4) &=& 1783.42519 + 1932.76417\, d_{44}^{FA} - 36.52029 \, d_{44}^{FF}\,, \\
\gamma^{(3)}_{{\mathrm{NS}}}(2,n_f=5) &=& -778.54684 + 1932.76417\, d_{44}^{FA} - 45.65037 \, d_{44}^{FF}\,, \\
\gamma^{(3)}_{{\mathrm{NS}}}(2,n_f=6) &=& -3047.70285 + 1932.76417\, d_{44}^{FA} - 54.78044 \, d_{44}^{FF}\,,
\end{eqnarray}
where $d_{44}^{FF}$ and $d_{44}^{FA}$ are the contributions coming from (\ref{d44FF}) and (\ref{d44FA}) correspondingly.

 \subsection*{Acknowledgments}
We would like to thank K.G. Chetyrkin, A.L.~Kataev, L.N. Lipatov, A.I. Onishchenko, A.V. Smirnov and V.A.
Smirnov for useful discussions.
This work is supported by RFBR grants 10-02-01338-a, RSGSS-65751.2010.2.

\end{document}